\documentclass[manuscript,nonacm]{acmart}

\usepackage{multirow}
\usepackage{booktabs}
\usepackage{xcolor,colortbl}
\usepackage{graphicx}
\usepackage{hyperref}
\usepackage{balance}
\usepackage[frozencache,cachedir=.]{minted}
\usepackage[position=top]{subfig}
\usepackage{tikz}
\usepackage{threeparttable}
\usepackage{nicematrix}
\usepackage{enumitem}
\usepackage{pdfpages}
\usepackage{multicol}
\usepackage{placeins}
\newlist{multiselect}{itemize}{2}
\setlist[multiselect]{label=$\square$}

\AtBeginDocument{%
  \providecommand\BibTeX{{%
    \normalfont B\kern-0.5em{\scshape i\kern-0.25em b}\kern-0.8em\TeX}}}

\definecolor{codebg}{RGB}{255, 255, 230}
\newcommand\encircle[1]{%
  \tikz[baseline=(X.base)] 
    \node (X) [draw, shape=circle, inner sep=0, fill=black, text=white] {\strut #1};%
}
\newcommand\squarenode[1]{%
  \tikz[baseline=(X.base)] 
    \node (X) [draw, shape=rectangle, inner sep=0, very thick, minimum size=3mm, fill=none, text=black] {\strut #1};%
}
\newcommand{\vrulesep}{\unskip\ \vrule width 2pt\ }

\begin{document}

\date{}

\title{"Sign in with ... \textit{Privacy}": Timely Disclosure of Privacy Differences among Web SSO Login Options}
\titlenote{\textbf{To appear, accepted on 18-Dec-2024 for publication in ACM Transactions on Privacy and Security (TOPS).}}

\author{Srivathsan G. Morkonda}
\affiliation{%
  \institution{Carleton University}
  \city{Ottawa}
  \country{Canada}}

\author{Sonia Chiasson}
\affiliation{%
  \institution{Carleton University}
  \city{Ottawa}
  \country{Canada}}

\author{Paul C. van Oorschot}
\affiliation{%
  \institution{Carleton University}
  \city{Ottawa}
  \country{Canada}}

\authorsaddresses{\textit{Email addresses: \href{mailto:srivathsanmorkonda@cmail.carleton.ca}{srivathsanmorkonda@cmail.carleton.ca} (Corresponding author), \href{mailto:chiasson@scs.carleton.ca}{chiasson@scs.carleton.ca}, \href{mailto:paulv@scs.carleton.ca}{paulv@scs.carleton.ca}}}

\renewcommand{\shortauthors}{Morkonda, Chiasson, and van Oorschot}

\begin{abstract}
The number of login options on web sites has increased since the introduction of web single sign-on (SSO) protocols.
Web SSO services allow users to grant web sites or \textit{relying parties} (RPs) access to their personal profile information from \textit{identity provider} (IdP) accounts.
Many RP sites fail to provide sufficient privacy-related information to allow users to make informed login decisions.
Moreover, privacy differences in permission requests across login options are largely hidden from users and are time-consuming to manually extract and compare.
In this paper, we present an empirical analysis of popular RP implementations supporting three major IdP login options (Facebook, Google, and Apple) and categorize RPs in the top 500 sites into four client-side code patterns.
Informed by these RP patterns, we design and implement SSOPrivateEye (SPEye), a browser extension prototype that extracts and displays to users permission request information from SSO login options in RPs covering the three IdPs.
\end{abstract}



\maketitle

\section{Introduction}
Web Single Sign-On (SSO) systems are widely used for user authentication, including by many popular web applications. For example, a user on \url{Airbnb.com} selecting ``Sign in with Google'' is redirected to Google's login page where they are prompted to log in using their Google account credentials; Google then verifies and conveys the user's identity to the \url{Airbnb.com} site.
More generally, an SSO system consists of a collection of entities that use a standard SSO protocol to enable users to authenticate to one or more relying party (RP) applications (e.g., sites like \url{Airbnb.com}) using an account registered with an identity provider (IdP, e.g., Google).

This is convenient for users as they can log in to different RP sites using a single IdP account which requires managing only a single set of credentials.
OAuth 2.0~\cite{oauth2rfc} is a standard authorization protocol that is widely used in SSO login systems to enable users to authenticate and grant RPs access to personal information (e.g., their full name, their calendars) from their IdP accounts.
Access to user data is provided by IdPs through OAuth-backed APIs that enable RPs to programmatically access a set of user data resources approved by the user during login.
RPs can access user data from an IdP to offer additional functionality (e.g., a personal calendar manager).

These APIs raise privacy concerns~\cite{balash2022security} as RPs could gain access to extensive user data (potentially built up by IdP account use over several years) typically without fully explaining or justifying the requested access to users.
The API access is not typically limited to the time of login or to a user session in RP site; it can include user data from the past, and can extend into silent continuous monitoring of future activities.

To engage a wider set of users, sites often offer SSO login options from more than one identity provider. The choice of login option can impact user privacy differently depending on three primary factors.
First, each provider exposes APIs giving access to user data that are relevant to their own app ecosystem. For example, Google offers the Gmail API for email retrieval by RPs; Facebook offers APIs for RP apps to access a user's photos and videos.
Second, API access is granted for broadly specified user data types and the API does not restrict access to data from a specific time range. This means that the amount of user data an API releases to the site depends on user factors such as how long ago the user's account was created and how much user activity is associated with the account.
Third, user privacy depends on the actual user data APIs the RP is designed to access through the SSO option selected by the user.

The number of login choices offered by RP sites has steadily increased over the past decade~\cite{jarpehult2022longitudinal}.
Most sites that offer multiple SSO login options request different types of personal data from each SSO, often with one option more privacy-friendly than others~\cite{morkonda2021empirical}.
It is also observed that RPs offer fewer but more privacy-friendly SSO login options to users in the EU compared to non-EU users, likely due to stricter privacy laws such as the GDPR~\cite{morkonda2021empirical}.
These privacy differences are often not visible as current SSO user interfaces (UIs) are designed to only inform the user after they make a choice (i.e., enter their credentials) of an SSO login option.
Current UI workflows also do not support easy comparison of requested data between the login choices offered by RP sites.
In cases where there are multiple alternate (especially more privacy-friendly) SSO login options, users might make choices that reveal more information than desired.
Moreover, a manual comparison requires the user to first complete login with each SSO option before seeing the requested permissions.

To address these issues, we considered how to inform users about the SSO permissions requested by an RP site.
We built a browser extension SSOPrivateEye (SPEye), that lists all the data resources the RP requests through individual SSO login options before the user enters their credentials on the IdP login page.
In addition, for a subset of RP sites, this tool extracts the permissions for each SSO login option and displays for users a privacy comparison as the user navigates to the RP login page before committing to a login choice.
Our contributions are:
\begin{itemize}
    \item Using empirical analysis, we identify and explain four client-side code patterns used by popular RPs to implement SSO login options for desktop browsers. For each of the four patterns, we give strategies allowing a tool to automatically extract RP site data useful for privacy and security analysis.

    \item We design and build SPEye, a browser extension displaying real-time privacy information about SSO permission requests. The information is presented in a standard format, just-in-time (before the desktop user decides to log in to an RP site).
    \begin{itemize}
        \item The approach takes into account privacy differences in permissions prompted to users in locations where different privacy laws apply.
        
        \item For one code pattern, SPEye also generates a privacy comparison of each SSO login option at the RP login page, reflecting the current RP site APIs that extract user information from IdPs.
    \end{itemize}

    \item We demonstrate the usefulness of our prototype\footnote{Source code available at: \url{https://github.com/choruslab/SSOPrivateEye}} by building it to recognize SSO login options with three major IdPs: Facebook, Google, and Apple. We designed the tool such that further IdPs (that use standard SSO protocols) can be added without major structural changes.
    
\end{itemize}

\section{Background and motivation}
Our privacy tool extracts information on SSO user data a given RP intends to access through each SSO login option listed on the site.
Here we summarize OAuth-based SSO protocols and present the privacy issues that motivate our work.

\subsection{SSO protocol background}
\label{sec.background.protocol}
SSO protocols commonly supported by major IdPs (e.g., Google, Facebook, and Apple) include OAuth-based protocols.
The OAuth 2.0~\cite{oauth2rfc} authorization protocol is designed to give an RP delegated access (within a specified scope) to user data in an IdP account. OpenID Connect 1.0~\cite{openIdConnect1Spec} is an adaptation of OAuth 2.0 designed specifically for user authentication. Although they are different specifications, they are closely related and both are offered by major IdPs.
\newline

\begin{figure}[tb]
    \includegraphics{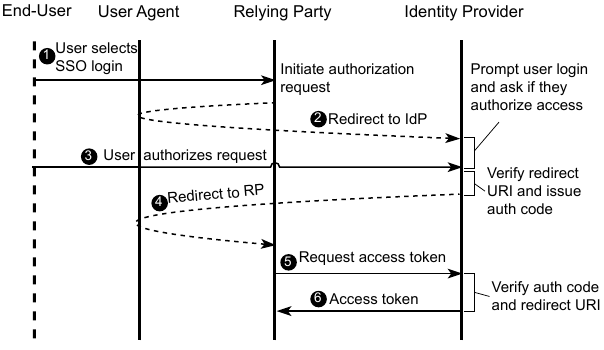}
    \caption{Overview of OAuth 2.0 authorization code flow.}
    \label{figAuthCodeFlow}
\end{figure}

\noindent\textbf{OAuth 2.0.}
OAuth 2.0~\cite{oauth2rfc} is an authorization protocol for granting RPs access to user resources protected by an IdP without disclosing the user's credentials to the RP.
This is achieved through \textit{access tokens} which are confidential strings issued by an IdP to allow the RP to access protected resources. Access tokens are obtained by executing one of several authorization procedures called OAuth \textit{flows} or \textit{grant types}. OAuth 2.0 has four main flows designed to support authorization from different resource owners such as SSO users and app services. RP sites that use OAuth for user SSO login primarily follow the \textit{authorization code flow} or the \textit{implicit flow}.
Although the two flows differ in security properties~\cite{oauth2rfc}, extracting protocol data from these flows involves an identical approach for our work.
For this reason, we describe only the authorization code flow (Fig.~\ref{figAuthCodeFlow}) in more detail and highlight shared attributes of these flows relevant to our work.

In Step \encircle{1}, the user selects SSO login at the RP site which initiates the flow by sending an \textit{authorization request}. This includes query parameters that inform the IdP about the RP's request to access user data. The \texttt{scope} parameter specifies the data resources (i.e., IdP APIs) the RP wants to access. Other relevant parameters include a client ID (unique string issued to the RP during a registration phase with the IdP), a grant type (to indicate OAuth flow), and a redirect URI to specify the endpoint to which the IdP should redirect the user agent once the request has been authorized (or denied) by the user.

In Step \encircle{2}, the RP redirects the user agent to the IdP's \textit{authorization server} URL where the user is prompted to login with their IdP credentials. Then, the user is asked if they want to grant RP access to the requested data (Step \encircle{3}). Once the user completes login at the IdP (with or without granting access), the IdP redirects the user agent back to the RP along with a fresh \textit{authorization code} value (Step \encircle{4}). Before redirection to the RP, the IdP must verify that the redirection URI specified in the SSO request matches a value provided by the RP during its registration with the IdP. This ensures that the authorization codes are delivered to the correct RP URL.

As a final step, the RP needs to exchange the authorization code for an access token. In Step \encircle{5}, the RP sends the authorization code (along with an optional client password for RP authentication) to the IdP's \textit{token exchange} endpoint where the request is verified. If the authorization code (and the client password) is valid, the IdP issues a fresh access token and returns it to the RP (Step \encircle{6}). The access token grants the RP access to data resources approved by the user.
It is also possible for the RP to obtain fresh access tokens and extend previously granted access by exchanging a \textit{refresh token} without involving the user for each new exchange.
\newline

\noindent\textbf{OpenID Connect 1.0.}
The OAuth protocol was originally designed for authorization, but through custom changes RPs could use OAuth to perform user authentication in SSO login. The OpenID Connect 1.0~\cite{openIdConnect1Spec} specification (OIDC) was developed as an authentication extension to OAuth 2.0. It introduces the \textit{ID token}, a value issued by an IdP (also referred to as an OpenID Provider or OP) to convey information about a user's identity to the RP.
Common data returned in an ID token include a unique identifier for the user, an expiry time for the token, and an IdP identifier.
A standard ID token uses the JSON Web Token (JWT)~\cite{jwtrfc} data structure to represent a digitally signed JSON object that contains verifiable \textit{claims} about an authenticated user.
Standard claims about the user include basic profile info such as name, email address, and profile picture.
ID tokens enable the RP to verify a user's identity based on the claims returned by the IdP.
The claims contained in the ID tokens could reveal sensitive personal information depending on specific IdP implementations.

OIDC extends OAuth 2.0 to define a set of six OIDC flows for RPs to obtain ID tokens, and access tokens when requested. Similar to OAuth 2.0, each OIDC flow begins with an initial \textit{authentication request} from RP to IdP with the \texttt{response\_type} parameter indicating the OIDC flow type.
Authentication requests use the same parameters as authorization requests in OAuth 2.0 but with special values. For example, the OIDC specification uses the \texttt{scope} parameter with the value ``openid'' to indicate an authentication request. Each flow then redirects the user agent to the IdP's login page where the user is prompted to grant the RP access to basic information such as their name and email address. If the user consents, the IdP then redirects the user agent back to the RP (at the specified redirect URI). Depending on the OIDC flow, the IdP might redirect to the RP with an authorization code (e.g., as in the authorization code flow) or directly return the ID token (e.g., in the implicit flow).

\begin{figure*}[tb]
 \centering
    \subfloat[Rakuten.com (RP) login]{
        \label{figRakutenLogin}
        \includegraphics[width=0.25\linewidth]{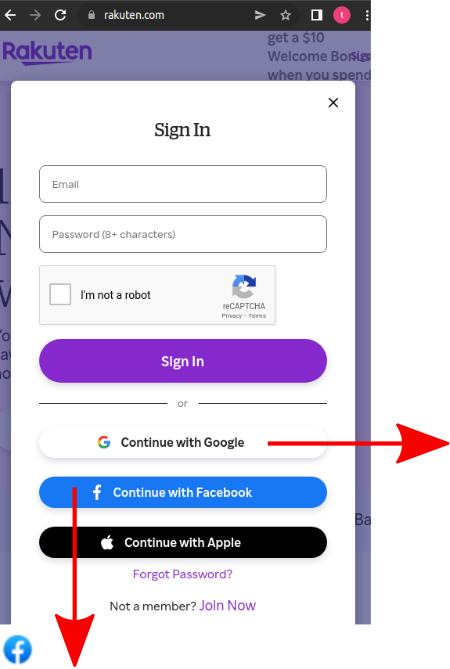}}
    \vrulesep
    \subfloat[Google (IdP) authorization form]{
        \label{figRakutenGoogle}
        \includegraphics[width=0.55\linewidth]{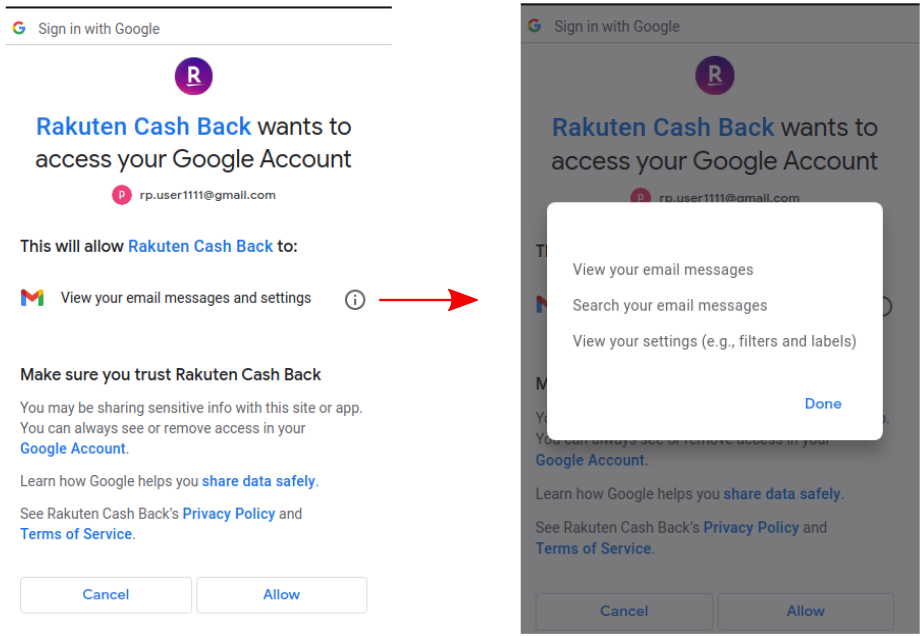}}
    \par\noindent\rule{\textwidth}{2pt}
    \subfloat[Facebook (IdP) authorization form]{
    \centering
        \label{figRakutenFB}
        \includegraphics[width=0.55\linewidth]{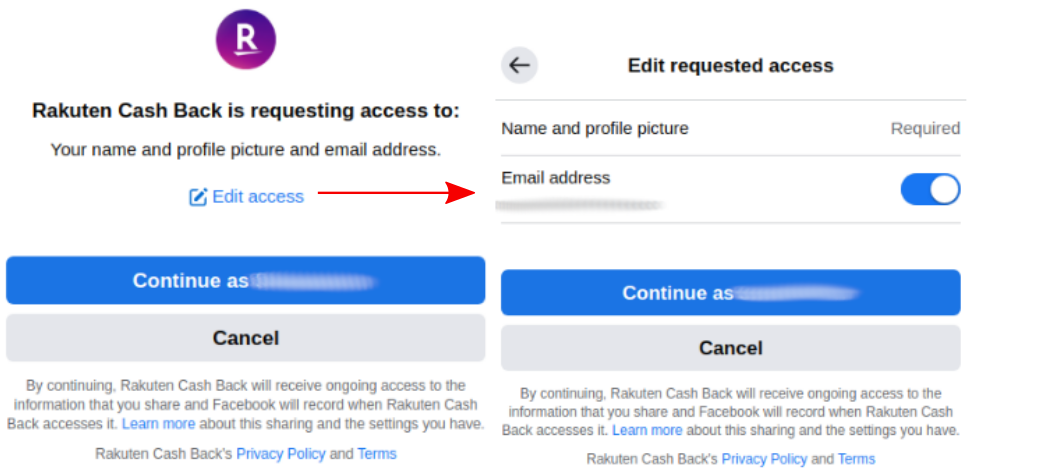}}
    \vrulesep
    \subfloat[Insufficient permissions error]{
        \label{figRakutenError}
        \includegraphics[width=0.25\linewidth]{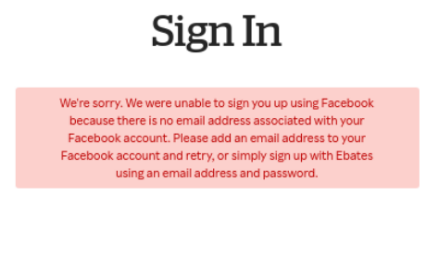}}
    \caption{Lack of transparent information at time of (a) user login prompt. When signing into Rakuten.com with (b) Google or (c) Facebook, the user is informed about permission requests only for the selected choice (typically after the user has committed to using that SSO). In (d), an attempt to login using Facebook without revealing the email address raises an insufficient permissions error on the RP site. Note the lack of justification on why data access is needed; to access this information, a user would need to navigate to and search a secondary page such as the RP's privacy policy.}
    \label{figRakuten}
\end{figure*}

\subsection{Privacy issues in current SSO UI design}
\label{sec.privacy.issues}
Our work aims to address SSO design-related privacy issues associated with RP requests to access IdP user data.
In the discussion below, we first describe the typical UI design used in SSO procedures to inform users about permission requests. We then highlight motivating privacy issues in this design.

As observed in recent studies (e.g.,~\cite{jarpehult2022longitudinal, morkonda2021empirical, jannett2024ssomonitor}), many RPs offer multiple SSO options to support users across popular IdPs.
User data accessed through OAuth APIs is useful for RPs wanting to give a more personalized user experience and offer extended functionality.
For example, an RP site might autofill forms using personal information from the IdP or offer tools to edit photos in the user's IdP photo library.
In some cases, this API data is optional, meaning that the RP site can function as designed when a user authenticates using SSO login but denies access to a subset of requested data. In other sites, RP services such as sending emails on behalf of the user require access to the user's IdP data (e.g., email account). This distinction between the user's IdP data that is essential and data that is optional for provisioning RP services is often unclear based on information given to inform users. We illustrate this with the sample SSO UI below.

A typical RP login form lists one or more SSO login choices along with traditional username and password fields for non-SSO (i.e., site-specific) accounts.
If a user picks an SSO login, their browser agent redirects to the IdP where they are prompted to enter their IdP account credentials (if they are not already signed in with that IdP). Then, the IdP informs the user of the data access requested by the RP.
Fig.~\ref{figRakutenLogin} highlights the UI design in a typical SSO login flow from the RP to the individual IdPs.
For SSO choices in Fig.~\ref{figRakutenLogin}, the figure illustrates the UI prompts (Fig.~\ref{figRakutenGoogle} and ~\ref{figRakutenFB}) where information on permission requests is displayed.
This design raises two main questions about the transparency of permission requests:

\begin{itemize}
    \item \textit{Is the access necessary to use the service?} Both the RP and IdP UIs lack information on whether the requested access is essential for the RP to offer its services. For example, the Google login dialog in Fig.~\ref{figRakutenGoogle} informs users that \texttt{Rakuten.com} wants access to the user's email messages but it is not clear whether this data is required for the site to function properly. Note that it is also possible to use the site with the Facebook (or the Apple) login option which only require the user to disclose their name and email address (Fig.~\ref{figRakutenFB}). The secondary prompt in the Facebook login UI (in Fig.~\ref{figRakutenGoogle}) suggests an opt-out option for the email address. However, we find that an attempt to login without revealing the email address raises an error (Fig.~\ref{figRakutenError}) on the RP site asking the user to retry using their email address.
    \item \textit{Which login option is more privacy friendly?} The second issue is the lack of visibility across the available SSO choices. The OAuth APIs exposed by an IdP depend on the type of user data it stores and the IdP's privacy policy for sharing user data with RPs.
    Users may not be aware of data requested by other SSO choices as they are only informed (after completing authentication) about the permissions requested with the SSO option they select.
    If a user is aware that an alternate SSO choice reveals less personal data compared to other options, they might choose differently. In current SSO UI design, the user would need to login with each SSO choice and manually compare the permissions to be fully informed about the privacy choices. This is time consuming as it could require entering credentials (and completing two-factor verification where enabled) with each IdP login.
\end{itemize}
We highlight that at the key decision point of selecting an SSO option and entering their IdP credentials (which occurs in Fig.~\ref{figRakutenLogin}), the user lacks the knowledge necessary to make an informed comparison of options.
These privacy issues can lead to users making privacy decisions without full information.
Our work aims to address this through a browser extension that automatically extracts SSO authorization requests and reveals the permissions that would be requested by the RP after login.
It provides this information without the user completing login at each IdP; and for a subset of RPs, the extension generates a comparison of IdP permissions earlier on the RP login page, thus further reducing user effort required to compare available privacy choices and make informed login decisions.
Our tool addresses the second issue above by enabling users to identify which options are better aligned with their privacy preferences, thus offering better control over their privacy when using SSO to login.

Addressing the first issue is more challenging because it is difficult to predict at which point in its service workflow an RP might request extra permissions, as discussed in Sec.~\ref{sec.discussion.otherconsiderations}.
While our tool does not directly address this issue, we hope that our comparison before the login prompt increases the user's ability to make an informed choice.

\section{Analysis of client-side code patterns}
\label{sec.code.patterns}
OAuth 2.0~\cite{oauth2rfc} is a complex framework, consisting of various flows and mechanisms to support authorization.
The design of the  OAuth 2.0 framework has motivated several OAuth 2.0 extensions (in separate RFCs) that target a multitude of use cases and authorization mechanisms~\cite{parecki2019time}; as of May 2024, there are 30 OAuth 2.0-related documents produced by the IETF (Internet Engineering Task Force) Working Group for OAuth.\footnote{\url{https://datatracker.ietf.org/wg/oauth/}}
Many of the implementation details are left to the RPs and IdPs, including how to implement authorization requests made by RP websites.

RP implementations of OAuth and OIDC protocols differ significantly across RP sites, and can include variations within an RP site for the implementation of individual IdP login options.
For example, some RPs use an IdP-provided SDK to exchange protocol messages with the IdP while others use front-end RP code (HTML or JavaScript) or back-end RP server code (such as Java or PHP) to manage the OAuth flows.
These variations complicate automated security and privacy analysis of OAuth systems, and a single approach may not be sufficient to cover or collect data from every RP site, as exhibited by previous studies~\cite{drakonakis2020cookie,ghasemisharif2018single,zhou2014ssoscan}.
Thus, it is essential to understand the main approaches by which RP sites implement authorization requests in order to extract information on SSO permissions.
The extracted information can be used to help inform users about the privacy implications discussed in Section~\ref{sec.privacy.issues}.

\subsection{Analysis Methodology}
Authorization requests are triggered by the user, typically by clicking a button or link from the RP site which redirects to the IdP with several protocol parameters (Sec.~\ref{sec.background.protocol}) relevant for security and privacy.
To guide the design of our SPEye tool (Sec.~\ref{sec.tool}), we examined the implementations of RP authorization (in OAuth 2.0) and authentication (in OIDC) requests\footnote{For simplicity, we refer to requests in OAuth 2.0/OIDC as SSO requests.} in the top 500 sites of the Tranco~\cite{lepochat2019tranco} list. We focused our empirical analysis on the RPs that implemented login with Facebook, Google, and Apple, these being the most common IdPs~\cite{morkonda2021empirical}.
We found 153 RPs that used SSO services of these IdPs.
We analyzed the SSO requests implemented by these RPs using the following steps:
\begin{enumerate}
    \item First, the lead researcher used an off-the-shelf browser to manually visit each of the top 500 websites to identify sites that offered SSO login options with Facebook, Google, or Apple. This resulted in a dataset of 153 RPs.
    \item For each RP, we analyzed the website's implementation of SSO with each individual IdPs. The goal was to identify the approach used by the RP to make SSO requests. We systematically explored the login page's HTML and JS code where these requests are handled after a user triggers login:
    \begin{enumerate}
        \item Using the browser's DevTools webpage inspect function,\footnote{\url{https://developer.chrome.com/docs/devtools}} we manually searched the HTML code that renders the SSO login options for embedded attributes (such as \texttt{href} links and form \texttt{action}) linked to SSO requests. We identified relevant attributes by checking if OAuth (or OIDC) parameters were included, and by checking for IdP APIs and endpoints.
        \item In the case where no relevant HTML attribute was found, we searched the RP site's JS files for code related to SSO requests, including event listeners attached to the UI components. We also found related code by searching the JS files for common keywords such as ``oauth'', ``openid'', and ``SSO''.
        \item For each RP, we monitored the browser's DevTools network log of the traffic generated by the website after sequentially selecting each SSO login option. In particular, we focused on the chain of HTTP requests generated after the login trigger (when an SSO request begins) up to the IdP login prompt (when an SSO request completes).  While some sites may use code obfuscation to hide certain implementation details, the traffic requests and responses captured in the browser offer insights about the implementation such as the use of IdP SDKs, intermediate RP services, and direct connections with IdP servers.
    \end{enumerate}
    \item After tracing each SSO login option from the UI components to the generated HTTP traffic, we categorized each login option based on the approach used by the RP to make SSO requests; these categories are discussed in Section~\ref{sec.codepatterns.results}.
\end{enumerate}

\subsection{Code Pattern Results}
\label{sec.codepatterns.results}
We identified four client-side code patterns (i.e., implementation approaches) for RP implementations of SSO requests, and counted how often each occurred: (i) 36 implementations used an HTML-based pattern; (ii) 56 used a JavaScript-based pattern; (iii) 44 used an IdP SDK; and (iv) 17 used a combination of the three patterns.
Each code pattern groups RP implementations using a similar coding approach for implementing SSO requests.
The HTML-based and JavaScript-based patterns include RPs that used HTML or JavaScript code for SSO implementations, respectively. The IdP SDK-based pattern includes RPs that used the IdP's recommended approach.
These distinct code patterns describe the various coding approaches used by RPs to request SSO permissions, and thus they help inform the design of the SPEye browser extension (Sec.~\ref{sec.tool}) to extract and display these permissions to the user.

Next we describe each pattern in more detail and discuss ideas for automatically extracting protocol data from RP sites.
Listing~\ref{code.patterns.examples} includes code samples for each of the three main code patterns described next.
\newline

\begin{listing}[tb]
\begin{minted}[bgcolor=codebg,frame=single]{html}
<!-- HTML pattern -->
<div class="sso-logins">
  <a id="sso-google" href="https://example.com/sso-google">
    <div>Sign in with Google</div>
  </a>
</div>

<!-- JS pattern -->
<div class="sso-logins">
  <button id="sso-fb" value="Login with Facebook"
  onclick="sso()"></button>
</div>
<script> function sso() {
  req = new XMLHttpRequest();
  req.open("POST", "https://example.com/sso")
  req.send('ssoWith=facebook');
 }
</script>

 <!-- IdP SDK pattern -->
<div class="sso-logins">
  <button id="sso-fb" value="Login with Facebook"
  onclick="sso()"></button>
  <script> function ssoFB() {
      FB.login(function(response) {
        // handler function for IdP response
      }, {scope: 'user_friends,user_likes'});
    }
  </script>
</div>
<script src="https://connect.facebook.net/en_US/sdk.js">
</script>
\end{minted}
\caption{Implementation of an SSO request in three different code patterns.}
\label{code.patterns.examples}
\end{listing}

\noindent
\textbf{1) HTML-based SSO.} In this code pattern, the SSO requests are embedded directly into the SSO-related HTML elements.
When the user selects an SSO login option, a request is sent to the link in the element's \texttt{href} attribute.
In most sites, we observed that this link leads to the RP server code which responded with an HTTP 302 code to redirect the user agent to the IdP endpoint. A small number of sites included the IdP link and the OAuth parameters directly in HTML; to avoid vulnerabilities, the OAuth \texttt{state} parameter must not be reused across requests (it should be a non-guessable nonce, to protect against CSRF attacks~\cite{oauth2rfc}).

In our dataset of 153 RPs, we found 36 implementations with SSO requests in the HTML code.
One RP, \texttt{ok.ru}, directly included the IdP endpoint in HTML; all other RPs implemented SSO requests through redirects from their backend servers.
These requests can be extracted by finding and sending GET requests to the RP. For each valid request, the RP server returns a response to redirect the user agent to the IdP endpoint. The OAuth protocol parameters are included in the redirection URL as query strings. We distinguished requests to endpoints of each targeted IdP using a set of URLs associated with each IdP.

We identified implementations with this code pattern by scanning the HTML of RP login pages for \texttt{href} and \texttt{form} tags, including endpoints to known RP and IdP servers.
\newline

\noindent
\textbf{2) JavaScript-based SSO.}
In this code pattern, DOM events such as button clicks on RP login page trigger RP JavaScript code and generate SSO requests.
We found that most sites send the SSO requests to their backend servers before redirecting to the IdP endpoints. 
This design is useful when the RP wants to initialize and maintain per-user state information (separate from the OAuth \texttt{state} parameter) in its backend server.
If the RP functionality is fully front-end implemented, this implies that the authorization request is implemented entirely in JavaScript with no backend server communications.
The OAuth \texttt{state} parameter (returned by the IdP after login) allows the RP's callback code to link the IdP response to the corresponding SSO request.
Scanning implementations of this pattern by analyzing (but not executing) loaded HTML page elements is challenging as RP JavaScript code often includes parameters such as \texttt{scope} evaluated only during runtime.

We found that 56 of 153 RPs sent the SSO requests from JavaScript code.
Although we did not exhaustively search every RP script (from code visible at the client-end), we found many sites that did not include authorization parameters in JavaScript code.
Instead the parameters are located in server responses to dynamically constructed HTTP requests.
In these implementations, extracting information (including protocol parameters) related to authorization requests may not be possible solely by searching code visible at the client (e.g., searching for pattern matches to static strings);
rather it might require dynamically executing RP code to simulate user clicks.
However, this is not suitable in tools for users (such as ours) as it may involve opening new browser windows and increase latency.

To identify RPs with this code pattern, we monitored the network traffic after manually clicking each SSO button on the RP's login page. If a request was sent to the RP server (and if the link was not embedded in HTML), we searched the JavaScript in RP HTML pages for traces of the request URL. This pattern can also be identified by searching for listener functions (e.g., specified in \texttt{onclick} attributes) that are triggered when an SSO-related element is clicked.
\newline

\noindent
\textbf{3) IdP's SDK-based SSO.}
Some IdPs provide software development kits (SDKs) for RPs to integrate their SSO services.
RP sites use these SDKs by importing the SDK library to manage SSO requests and responses with the IdP.
Although these libraries are designed to make it easier to integrate IdP services, these SDKs often make implicit security assumptions that might not be well understood by developers~\cite{wang2013explicating}.
These libraries also contain functions that allow the IdP to provide a consistent user experience across different RP and IdP sites. For example, the Google Sign-In library for JavaScript apps~\cite{googleApiAuth} offers the ``One Tap'' feature which allows the RP's landing page to include Google's sign-in prompt for asking the user to log in using Google.

We found that 44 of 153 RPs used IdP SDKs to manage authorization requests.
Unlike the other code patterns, all these SDK-using sites send requests directly to the IdP.
Therefore, the authorization request parameters (specified as arguments to the SDK functions) are available in RP's JavaScript code.
Analyzing the RP's client-side code could help to identify SDK function calls.
Searching and extracting these parameters might be simple if the arguments are included directly in the function calls.
However, if the arguments are formed by combining other variables, extracting them might require more advanced methods such as data-flow analysis.

These implementations can be identified by searching for the presence of IdP libraries which are typically imported into HTML using \texttt{script} tags.

\textit{\textbf{IdP SDK privacy concerns.}}
When a user visits an RP site that imports multiple IdP SDK scripts (typically embedded in the landing page), a request is made to each IdP.
This allows the IdPs to track a user's RP visits, including a user who might hope to avoid IdP tracking of their browsing activity by choosing non-SSO login.
In the other code patterns (where IdP scripts are absent), only the user's chosen IdP learns about the user's RP visit, instead of multiple IdPs.
Sec.~\ref{sec.discussion.otherconsiderations} provides further discussion on third parties.
\newline

\noindent
\textbf{4) Mixed SSO.}
We found 17 RPs that implemented the SSO requests using more than one pattern.
We observed that \texttt{etsy.com} implemented SSO with Google and Facebook using IdP SDKs. Although Apple offers its own SDK~\cite{appleSignInDocumentation}, the site had implemented Apple SSO by including the IdP URL directly in its HTML.
We are not sure why sites implemented SSO options using different patterns. Perhaps some developers (adding a new IdP option) simply prefer a different code pattern than colleagues who implemented previous options.

\section{SSO-Private-Eye tool (SPEye)}
\label{sec.tool}
SPEye is our new prototype extension for the Chrome browser. It aims to inform users about the privacy implications of using SSO to log in to an RP site by enabling comparison of available SSO choices.
We present its design and implementation below.

\subsection{Design goals}
\label{sec.tool.requirements}
We designed SPEye to scan RP/IdP login pages (after page loading) for SSO services and generate permission information for available SSO login options.
We implemented SPEye as a browser extension due to three main design goals:
\newline

\noindent\textbf{G1) Real-time comparison.}
RPs often update the SSO services they offer. For example, after ``Sign in with Apple'' was introduced in 2019, Apple quickly overtook Twitter as the third most popular IdP~\cite{jarpehult2022longitudinal}. 
Tools that aim to inform SSO users about privacy practices should ideally use up-to-date data to take into account recent RP changes.
By extracting permissions at the time of login prompt, SPEye gives users up-to-the-minute privacy data of the available SSO choices.
\newline

\noindent\textbf{G2) User-location-specific comparison.}
Some RPs present different versions of their sites, e.g., to users within different countries, and request varying amounts of user information through the individual SSO login options~\cite{morkonda2021empirical}.
This practice might be a result of stricter privacy laws in some regions such as the EU's GDPR and the CCPA in the US.
To account for RP site differences, user tools need to consider the current location of the user.
SPEye uses local scripts within the user's browser application, and therefore extracts information from the RP site relative to the user's current location.
\newline

\noindent\textbf{G3) End-user focus.}
We designed SPEye to run its analysis without navigating the user away from the current RP or IdP login page to minimize disruption to their workflow.
Tools that are not targeted for end-users might not be subject to this constraint.
Several previous tools targeted for researchers (e.g.,~\cite{yang2016model, ghasemisharif2018single, morkonda2021empirical, jarpehult2022longitudinal, jannett2024ssomonitor}) have instead extracted the protocol data using browser automation tools (such as Selenium~\cite{seleniumWebDriver}) to open a scripted browser window and simulate user actions in the SSO workflow.
SPEye's constraints differ for two reasons: (a) our tool is meant for end-users, so disrupting the user's view away from the current login page (e.g., by opening new windows) would limit usability, and (b) our tool does not require executing the entire SSO workflow as it only needs to extract the authorization request to a specified IdP.
\newline

\noindent
\textbf{Centralized design alternative.}
A potential alternative proposal that we did not follow due to its limitations is to design the browser extension to query results data from a central database server populated by a backend tool (e.g., a crawler that visits and analyzes a list of RP sites).
As described in G1 and G2 above, RP requested permissions are frequently updated. Even with frequent crawls to provide up-to-date data: (a) changes across locations would require the crawler to rely on VPN services, which can be identified/blocked by RPs/IdPs~\cite{morkonda2021empirical}, and  (b) some RPs present different SSO features based on user data (e.g., GDPR-compliant site version for EU citizens)~\cite{morkonda2021empirical}.
Therefore, privacy tools that collect and analyze RP data from a central location would need user-specific data (e.g., citizenship as above) unavailable to a crawler.
SPEye's client-side design resolves these issues.

\subsection{Design of workflow modes}
\label{sec.tool.scope}
Privacy information on RP permissions could be presented to users at two points in a typical SSO login workflow: (1) on a given IdP login page \textit{before} the user enters their IdP credentials; (2) on the RP login page where all available login options might be listed. 
We consider both options:

\textbf{Focused mode (IdP login page).}
As the user navigates from an RP to an IdP login page (after selecting an SSO login option), SPEye extracts and displays permission information about the data access the RP intends to request from this IdP.

\textbf{Comparative mode (RP login page).}
For a subset of RPs, a second mode additionally offers a comparison of permissions requested by the RP for each SSO login option while the user is still on the RP login page (before selecting any login option). In Sec.~\ref{sec.code.patterns}, we identified four RP client-side patterns;
SPEye's Comparative mode is currently available for the HTML-based pattern to extract comparative information about privacy (extendable to security) implications of SSO services for display to users on the RP login page.

\subsection{Design architecture}
\label{sec.tool.implementation}
The Focused mode of SPEye involves IdP login pages. These IdP login pages and their endpoints remain more consistent compared to RP login pages, whose SSO characteristics vary considerably from one site to another (as described in Sec.~\ref{sec.code.patterns}), thus leading to a more robust Focused mode implementation.
Implementation of the Comparative mode (on RP login pages) is more complex as it relies on heuristics to automatically identify SSO login options and extract permission information, based on earlier code analysis from different RP implementations.
For SPEye's Comparative mode implementation, we used the 36 HTML-based RP sites (Sec.~\ref{sec.code.patterns}) to design and test our approach.
We tested SPEye on a VM running Ubuntu 22.04.2 and Chrome version 113.0.5672.

\subsubsection{Focused mode implementation}
SPEye's Focused mode extracts permission data on each of the three supported IdP's login pages. 
The permissions are displayed on a per IdP basis as the user visits each given IdP login page, and before entering IdP account credentials.
Compared to workflows without SPEye, this avoids the need to login with each IdP to compare permissions, and may also reduce the amount of information shared.

However, this introduces a usability penalty in that users need to click on an individual IdP login button to access details about that IdP. Given the challenges with implementing Comparative mode, discussed below, we viewed this as a reasonable compromise, given also that this task  need only be completed when a user is debating login options for an RP (as opposed to during every login).

\subsubsection{Comparative mode implementation}
\label{sec.tool.testing}
To identify RP web pages with SSO login options, SPEye's Comparative mode searches the DOM for SSO elements using a list of CSS Selectors.
A similar heuristics-based approach is used by other tools (e.g.,~\cite{zhou2014ssoscan}) to identify SSO login options.
We divided the 36 HTML-based RP sites into a \textit{training set} of 21 sites and a \textit{testing set} of 15 sites.
We used the \textit{training set} to build heuristics for identifying SSO login options (i.e., CSS Selectors --- see Appendix~\ref{sec.app.identifying.sso.components}), and to determine how to extract authorization requests for the identified login options.

Extracting protocol data from RP login pages (including the custom RP implementations described above) might be possible with dynamic approaches involving browser automation (e.g., simulating user clicks), but such tools introduce other challenges including undesired impact on usability, incomplete RP coverage, and increased performance overhead.
With our Comparative mode, we chose to focus on the subset of RP sites where we can avoid such undesirable effects and provide reliable comparisons.  

\begin{figure}[tb]
    \centering
    \includegraphics[height=0.43\linewidth]{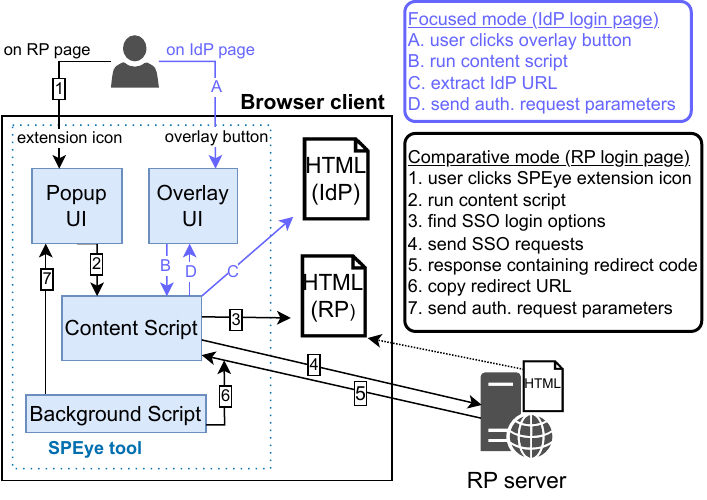}
    \caption{Architecture of SPEye and its two workflow modes. The Focused mode overlay button is displayed on every IdP login page, and the Comparative mode extension icon is available on a subset of RP login pages.}
    \label{figToolDesign}
\end{figure}

\subsection{Implementation of browser extension}
\label{sec.tool.extension}
Fig.~\ref{figToolDesign} shows the high-level interactions between SPEye components and the RP and IdP web site stack.
SPEye has four main components: a) a \textit{popup script} with \textit{popup UI} triggered on an RP login page when a user clicks on the \textit{SPEye extension icon} on the browser toolbar to open it; b) an \textit{overlay UI} on the IdP login page, opened through an \textit{overlay button} displayed by SPEye; c) a \textit{content script}, which runs on the visited HTML page when the interface is opened; and d) a \textit{background script} monitoring browser traffic for HTTP redirection requests to IdP endpoints.

SPEye scans the current page only when the extension interface is opened.
This prevents SPEye from interfering with user-triggered requests during which the interface remains closed.
It also eliminates unnecessary background processing of all the pages the user visits (while the extension is active) and limits performance impact.
SPEye interface triggers the following tasks (Step \textbf{A} or Step \squarenode{1} in Fig.~\ref{figToolDesign}).
\newline

\noindent
\textbf{1) Search RP/IdP HTML page.} 
Based on whether the user is on a login page of an IdP (Focused mode) or an RP (Comparative mode), the following steps are executed:
\newline

\begin{figure*}[tb]
 \centering
    \subfloat[Focused mode]{
    \label{figSpeyeToolPrimary}
        \includegraphics[width=0.3\textwidth]{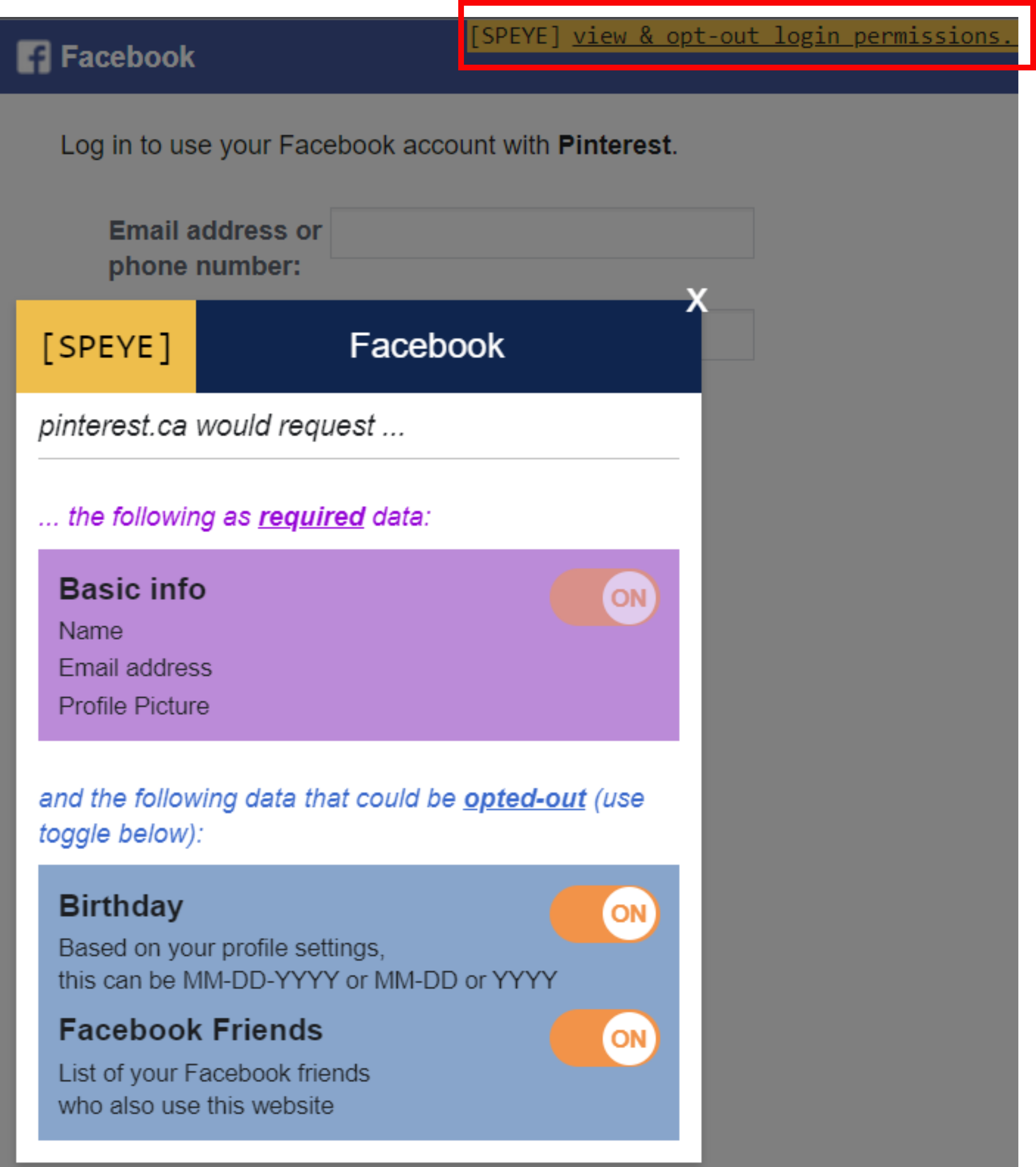}}
    \subfloat[Comparative mode]{
    \label{figSpeyeToolSecondary}
        \includegraphics[width=0.8\textwidth]{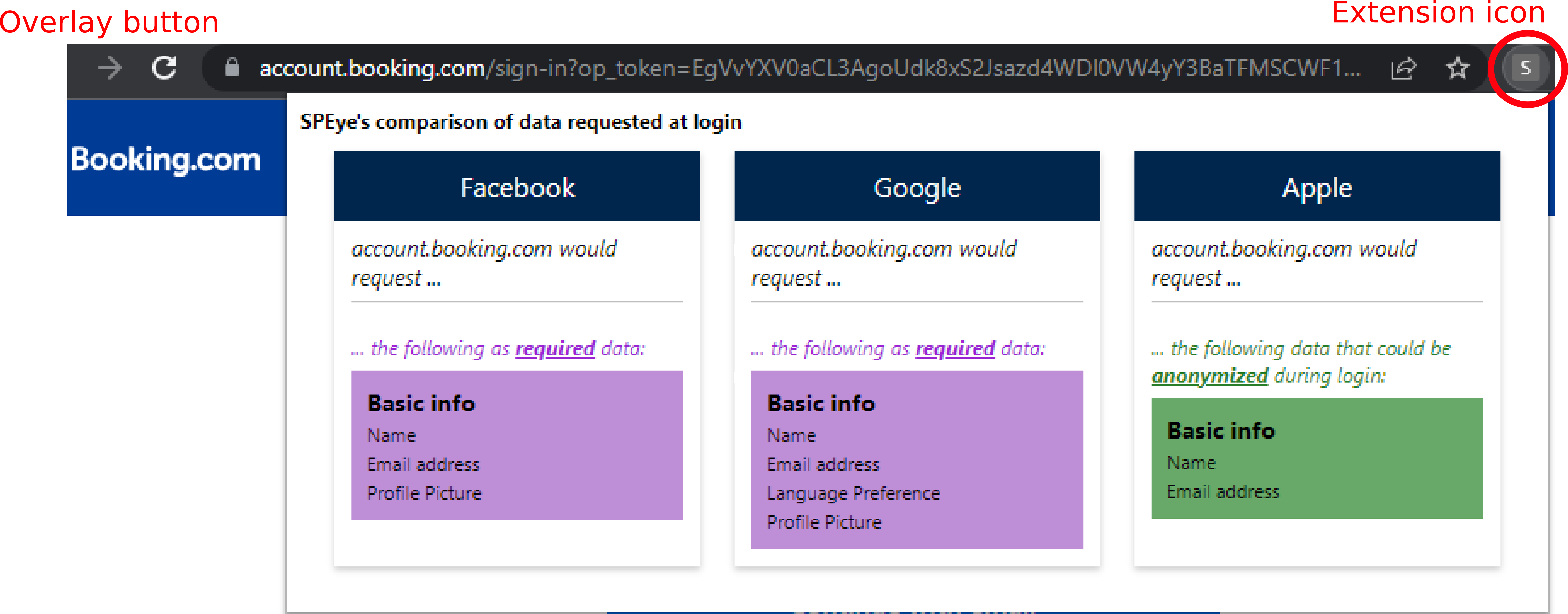}}
    \caption{SPEye's UI for (a) Focused and (b) Comparative workflow modes showing permission information on example sites.
    In the case of Apple SSO, SPEye also indicates Apple's privacy feature (not shown in this image but available on Apple's IdP UI) that allows the user to anonymize their information released to the RP.
    To get this view of permission data without SPEye, the user must login with each SSO option to manually collect, record, and compare the personal data requested.}
    \label{figSpeyeTool}
\end{figure*}

\noindent\textbf{1a) Focused mode.}
When the current (in-focus page) URL matches a known IdP authorization server, SPEye displays the overlay button (see top right in Fig.~\ref{figSpeyeToolPrimary}). The user can click (Step \textbf{A}) to view the permissions, and opt-out of certain permissions requested by the RP.
Using the Chrome runtime API,\footnote{\url{https://developer.chrome.com/docs/extensions/mv3/messaging/}} the interface script sends a message to the content script (Step \textbf{B}) to process the current page, extracting permission information from the RP's authorization request to the IdP.
\newline

\noindent\textbf{1b) Comparative mode.}
When the extension icon (Fig.~\ref{figSpeyeToolSecondary}) is clicked, the content script uses CSS Selector strings to identify RP login pages based on matching SSO buttons, login forms and login URLs (Step \squarenode{3}).
When a matching element is found, the script searches its attributes (e.g., \texttt{href}, \texttt{onclick}) for URLs to RP and IdP servers.
In the common match case (as observed in our dataset), the script extracts the URL from an \texttt{href} attribute and makes an \texttt{XMLHttpRequest} for each matching element (Step \squarenode{4}).

If a link is not found, the matching element could be part of a HTML form, identified by the presence of attribute types such as ``input'' and ``submit''.
For these matches, the script searches the DOM to identify its parent form element and extracts the path to which the form will be submitted along with other form data linked to a specific SSO choice.
As an example of form data observed in our dataset, selecting login with Facebook on \texttt{fandom.com} involved a custom form attribute \texttt{value="facebook"}.
SPEye's content script makes an \texttt{XMLHttpRequest} to submit the form with these custom attributes (sequentially, for each SSO login) along with other internal parameters (e.g., nonce values) in the form (Step \squarenode{4}).
\newline

\noindent
\textbf{2) Extract authorization request parameters.}
SSO design requires RPs to redirect the user to the IdP login page with authorization request parameters in the URL.
SPEye compares each redirection URL against a set of regular expressions (listed in Appendix~\ref{sec.app.identifying.sso.components}) derived from known IdP authorization server URLs.
\newline

\noindent\textbf{2a) Focused mode.}
If the user is on the IdP login page, SPEye extracts the parameters from the current IdP page without making any additional requests.
SPEye's content script compares the current URL (Step \textbf{C}) to match with a known IdP server.
Then, the script extracts the RP domain name and the list of permissions requested by the RP from the protocol parameters in the IdP URL.
The extracted results are sent to the overlay UI (Step \textbf{D}) for displaying to the user.
\newline

\noindent\textbf{2b) Comparative mode.}
If the user is on the RP page, SPEye's content script automatically triggers the SSO requests identified in the RP client-side code.
When the RP server accepts the request, it responds with an HTTP redirect code (Step \squarenode{5}).
SPEye's background script uses the Chrome webRequest API\footnote{\url{https://developer.chrome.com/docs/extensions/reference/webRequest/}} to observe HTTP redirects (Step \squarenode{6}).
If the redirection URL is identified as a known IdP server, the background script copies and sends the URL to the popup script for further processing (Step \squarenode{7}).
Although the extension's background scripts are always active (e.g., listening), redirect URLs are only received by the popup script when the SPEye user interface is opened by the user.
Other redirects such as user triggered SSO logins, during which the SPEye interface remains closed, are unaffected by the background script.

\textbf{\texttt{theguardian.com}} is an example that illustrates the type of typical variations we found across RP implementations.
We observed RPs that redirect through multiple endpoints before sending the user agent to the IdP server. In this example, the site's backend server responded to SPEye's SSO requests by returning a redirect to an intermediate (RP) endpoint which then returned with the IdP redirection URL.
In such instances, SPEye follows the sequence of redirects to observe the eventual IdP endpoint. 
\newline

\noindent
\textbf{3) Display permission information.}
SPEye aims to present the extracted permission information to the user in a consistent interface across different RP sites and IdP SSO platforms. 
We reviewed IdP documentation to map each OAuth \texttt{scope} parameter value to descriptive text explaining the associated permissions.
Where scope values from different IdPs refer to the same category of user information (e.g., user's IdP profile info), we modify the text provided by the IdPs to provide a consistent description.
\newline

\noindent\textbf{3a) Focused mode.}
On the IdP login page, SPEye users can view permissions requested by the RP as illustrated in Fig.~\ref{figSpeyeToolPrimary}.
This option is available before users enter their credentials, thus removing the effort of having to complete login with the IdP (along with multi-factor authentication, if applicable) before being able to view the RP requested permissions with that IdP.
This overlay UI also contains toggles to opt-out of certain permissions deemed optional by the IdP platform. When an opt-out is requested by the user, SPEye sends the IdP a new authorization request without the opted-out permissions.
The IdP might override the modified request and instead display a pre-registered set of permissions. 
The user may still be able to opt-out of some of these permissions through the IdP UI.
\newline

\noindent\textbf{3b) Comparative mode.}
On the RP login page, Comparative mode (Fig.~\ref{figSpeyeToolSecondary}) shows a privacy comparison of SSO login options.
The background script extracts the permissions from authorization requests (similar to the Focused mode above).
SPEye generates and displays a comparative summary of permission descriptions using the identified requests. 

Independent of the Comparative mode, SPEye users can visit an IdP login page to view the Focused mode output (in RPs that follow standard OAuth).

\subsection{Evaluation}
\label{sec.app.speye.evaluation}
As described in Section~\ref{sec.tool.testing}, we divided the 36 HTML-based SSO implementations used for SPEye's Comparative mode into a training set and a testing set.
In this section, we present the evaluation of SPEye's Comparative mode.
We note that the Focused mode achieved 100\% coverage in all RPs that follow standard OAuth (i.e., whether HTML, JS, SDK, or Mixed pattern SSO implementations).

\subsubsection{SPEye's Comparative mode misses}
\label{sec.app.speye.misses}
After implementation using the training set, the Comparative mode identified SSO login options and extracted authorization requests in 19 of 21 RPs (90\%) in the training set.
We then tested the Comparative mode implementation using the RPs in the testing set, and found extraction of comparative information in 12 of 15 RPs (80\%).
Looking across both the training and testing sets totalling 36 RP sites, all 5 misses were due to custom RP implementations that prevent automated tools like SPEye from extracting protocol data from RP login page.
We provide details on two different RPs to illustrate the Comparative mode misses:

\textbf{\texttt{unsplash.com.}}
Although SPEye's Comparative mode extracted and sent the SSO request to the correct URL, the RP server returned a non-redirect response.
Closer inspection of the site's HTML page reveals \texttt{<meta>} tags containing custom CSRF token parameters~\cite{owaspCSRFinDOM}.
Extracting these is beyond scope in the current version of SPEye as they might be added to the login requests from the RP's JavaScript code.
However, to confirm our findings, we modified SPEye to resubmit the request that included these parameters, and the RP server responded with a redirect request for the selected choice.

\textbf{\texttt{themeforest.net.}}
SPEye's requests to this RP's server were initially blocked.
To identify why, we compared a request sent by SPEye with the same request triggered through the SSO button on the RP page.
The requests were identical except for two HTTP request headers (\texttt{sec-fetch-dest} and \texttt{sec-fetch-mode}) typically added automatically by the browser. The values in these headers are used to inform the server that the request was initiated by a user clicking a link.
SPEye cannot modify these parameters as they are read-only,\footnote{\url{https://developer.mozilla.org/en-US/docs/Web/API/Request/mode}} and managed by the browser.

\subsubsection{IdP deviation from standard protocol}
\label{sec.tool.idp.deviation}
SPEye extracts RP requested permissions by inspecting the RP's authorization requests, as per standard OAuth protocol.
In February 2023, we observed that after manually signing in using Facebook to certain popular RP sites (e.g., \texttt{airbnb.com}, \texttt{cbc.ca}), the IdP prompt changed to show only the basic permissions (i.e., name and email address) and, contrary to expectations from standard OAuth, non-basic permissions such as photos were not displayed or granted.

To measure the extent of such anomalies, we first identified 133 RPs with Facebook SSO in the Tranco top 1000 sites and compared permissions in the authorization requests made by each RP to the permissions displayed in the IdP prompt (shown after login).
For 113 sites (85\%), we found that the permissions in the RP authorization requests were consistent with the permissions displayed on the IdP login prompt.
For 14 sites (10\%), Facebook did not show a login prompt, instead displaying an error stating that users can ``log in when the app is reactivated'' (Appendix~\ref{sec.app.fbchanges} shows an example).
For the remaining 6 sites (5\%), the non-basic permissions included in the RP's request were pruned by the IdP, and hence not displayed or granted.
These sites included \texttt{tripadvisor.com}, \texttt{theatlantic.com}, \texttt{coursera.org}, \texttt{airbnb.com}, \texttt{cbc.ca}, and \texttt{aliexpress.com}.

These deviations could be due to recent changes to Facebook's SSO policy~\cite{Fb2023NewRulesForAdvancedAccess} that restrict certain sites from access to non-basic data.
It is possible that the protocol deviations we observed in these RPs are temporary until the RP apps are reviewed by Facebook under its new SSO permissions policy.

The aim of SPEye is to extract and display permissions as requested by the RP, assuming that the OAuth protocol is followed.
If an IdP deviates from the protocol (as in Facebook above) by modifying the set of permissions requested by the RP, the information displayed by SPEye reflects what a user should see at the IdP interface in a compliant login flow.
However, as explained above, this may differ from what a non-compliant IdP actually shows.

\subsection{User Interface design}
SPEye's output (illustrated in Fig.~\ref{figSpeyeTool}) enables users to compare permissions before they commit to a login choice.
For each SSO login, the UI lists the information requested by the RP through the IdP's user data APIs.
For a subset of attributes, SPEye also describes what data (amount or type) will be released to the RP if the associated SSO choice is selected.
For example, if the user's IdP profile privacy settings is set to only reveal their age, the birthday API might only reveal the user's year of birth to the RP.
By enabling comparison across SSO choices, SPEye provides users with information to make informed decisions aligned with their privacy preferences.
This design can also inform users about specific privacy features. For example, Apple offers users the option to hide their email address (as indicated for the Apple SSO option in Fig.~\ref{figSpeyeToolSecondary}) by generating a unique per-RP email address which relays emails between the RP and the user's email inbox~\cite{appleHideMyEmail}.

\subsection{Impact of displaying IdP permission information on login decisions}
\label{sec.tool.userstudy.note}
A companion paper~\cite{morkonda2023influences} presents the results of an IRB-approved, 200-participant user study on the factors that influence login decisions in RP sites.
The goal of the study was to investigate the impact of showing IdP permissions on login decisions, for example, whether information extracted by SPEye about IdP permissions could be used to nudge users towards privacy-friendly options.
The user study complements the work in the current paper as here we present the technical details of extracting and displaying the permissions, rather than testing its impact on users.
For the convenience of the reader, this section provides a brief summary of the user study design and results.

\subsubsection{Study design}
Participants were shown screenshots representing the output of SPEye or an equivalent tool for each SSO login choice on an assigned RP site and prompted for their login decisions before and after viewing the screenshot of permissions requested by the RP with each IdP.
Along with each login decision, participants were asked to provide reasons for their decisions.
We then compared the first (pre-information, i.e., as shown in the current website interface) and the second (post-information) login decisions to test whether the screenshots influenced participants to make privacy-informed decisions.
To ensure the responses were valid, participants could not change their first login choice response after viewing the permission screenshots. 

The study involved four popular RP sites (\texttt{Airbnb.ca}, \texttt{CBC.ca}, \texttt{NYTimes.com}, and \texttt{Rakuten.com}) that offered different SSO and non-SSO login options involving a variety of IdP permissions and thus a range of privacy choices.
These RPs offered three SSO login options (Google, Facebook, and Apple) and at least one non-SSO login option (email address and phone number options on \texttt{Airbnb.ca}, and only email address on the other RPs).
The survey recruitment tool allocated exactly 50 participants per RP, and showed each participant only one of the four RP sites to avoid priming effects.
Participants were also asked basic questions about their web login experience, and questions about their use of accounts with popular IdPs (including Apple, Facebook, and Google) and related applications and/or services.
The participants completed the study in 6 minutes and 23 seconds on average, and they were compensated \textsterling 1.50 for their time.

\subsubsection{Pre- and Post-information results}
The majority of participants (90\% of 199 responses) reported that they had used web SSO in the past to log into at least one website. Among the 198 responses on IdP accounts, 96\% had a Google account, 80\% had a Facebook account, and 62\% had an Apple account. 
Among the first login choices (pre-information), 110 participants (55\%) chose an SSO login option, with Google login being more popular (selected by 89 participants) than Facebook (11 participants) or Apple (10 participants) login options.
Many participants preferred Google login because of convenience as they were already logged in with Google on other Google apps (such as Gmail).
90 participants (45\%) selected a non-SSO login option, and common reasons stated for their choice involved privacy concerns such as SSO linking of multiple RP accounts with an IdP and releasing personal data to RPs.

56 participants changed their login decisions after viewing the permission screenshots.
Among these participants, 86\% ($n=48$) changed to a more privacy-friendly login option.
The primary reason for the change given by these participants (in their open-ended responses) was privacy concerns about the permissions requested with their first login choice made apparent by the permission screenshots.
Our interpretation of these findings is that tools such as SPEye can influence users to make informed SSO login decisions and help privacy-conscious users reveal less personal information with RP sites.
See the companion paper~\cite{morkonda2023influences} for further results and study details.

\subsection{Deployability and maintainability}
Like any browser extension, SPEye may require updates over time, for example, to address changes in protocols and browser APIs.
OAuth 2.0 and OIDC are among the major SSO protocols currently used on the web~\cite{alaca2020comparative}; SPEye's compatibility with these protocols supports its use with little deployment and maintenance efforts since these are expected to remain relatively stable.
Future SSO protocols could deviate from the standard OAuth 2.0 design, which might impact SPEye and other OAuth-based SSO tools (e.g.,~\cite{jannett2024ssomonitor,li2019oauthguard,zhou2014ssoscan,mainka2017sok}).
While we cannot predict future designs, we hope that SPEye inspires future SSO protocols to be more secure and privacy friendly (cf.~Sec.~\ref{sec.recommendations}).
SPEye also relies on Chrome browser APIs (such as webRequest) to intercept HTTP requests and responses related to the SSO flows.
Changes to browser software and its APIs in future browser versions could disrupt browser extensions and require modifications.
However, we expect such impact on SPEye to be minimal since it relies on a small number of standard browser APIs that are supported across major browsers including Chrome and Firefox.
Given that SPEye retrieves the permission information directly from the RPs and IdPs with each request, there are no issues with staleness of the information (e.g., no need to update databases, to verify whether a user has changed locations, or to verify whether an RP has changed requested permissions), which simplifies maintenance and deployability.

\section{Limitations and future work}
Here we report limitations of the current version of SPEye and provide ideas for future improvements.
We also discuss a potential security extension.
\newline

\noindent\textbf{RP coverage of SPEye Comparative mode.}
As implemented, the main limitation of SPEye's Comparative mode is the lack of coverage beyond HTML-based SSO implementations. In addition, RP sites may change their SSO implementation to avoid detection by the Comparative mode (and other tools similar to SPEye). The Focused mode offers an improvement by covering all RPs (that use standard OAuth or OpenID Connect) at the expense of a usability cost requiring users to visit individual IdP login pages.
However, we believe that SPEye remains useful as it increases the transparency of permissions without requiring any changes from current RPs, IdPs, and the SSO protocols.
\newline

\noindent\textbf{UX testing and potential enhancements.}
The user study (Sec.~\ref{sec.tool.userstudy.note}) in the companion paper focused on investigating the impact of showing information about SSO login permissions on login decisions.
The results suggest that privacy information such as that extracted by SPEye informs users about SSO login permissions and results in many users changing their behaviour by choosing more privacy-friendly login options.
The study only evaluated the usefulness of permission-related information.
However, the lack of usability testing of SPEye where users directly interact with the tool is a limitation which we hope to address in future work.
Thus, a possible future research direction could be to test the usability of SPEye's user interfaces and refine the user experience of SPEye's different workflows.
For example, the study could compare users' privacy choices between SPEye's Focused mode and Comparative mode workflows.
The study could also explore user perceptions of SSO systems and their trust of tools such as SPEye.
\newline

\noindent\textbf{User data usage by RP.}
SPEye's extraction shows what data an RP requests, but cannot determine how the requested data is used by the RP. Users might want to weigh any asserted purposes before granting access to personal data. While an RP's privacy policy might offer information on intended use, it is difficult to ascertain actual usage (or even intent) without access to RP systems or developers.
\newline

\noindent
\textbf{Expanding IdP support in SPEye.}
SPEye currently covers the top three IdPs supported by RPs.
RPs might offer other IdP login options that request different data.
In our dataset of the 153 SSO-supporting RPs (in top 500 sites; Sec.~\ref{sec.code.patterns}), we found 25 unique IdPs.
At least 81 OAuth providers exist according to a Wikipedia IdP list~\cite{wikipediaOAuthProviders}.
SPEye's code is modular such that expanding the list of IdPs supported can be done with relatively little redesign, e.g., adding to a list of known IdPs by including data relevant to the new IdP such as authorization endpoints and permission information lists.
\newline

\noindent\textbf{SSO privacy in mobile apps.}
The OAuth 2.0 spec~\cite{oauth2rfc} originally targeted web apps and is less suited for implementation in mobile apps.
The mobile ecosystem is sufficiently different that it has its own issues separate from web implementations. 
Platform differences in mobile OSs make it challenging to securely implement OAuth 2.0 in mobile apps.
A study of mobile app RPs~\cite{chen2014oauth} found a number of vulnerable apps primarily caused by custom solutions for storing and delivering protocol secrets.
For example, due to an absence of secure redirection mechanism (in iOS and Android) from an IdP site on a mobile browser to an RP mobile app, access tokens could not be delivered safely.
Moreover, UI constraints in mobile apps require a novel approach to inform mobile users.
Our work analyzed web SSO implementations and built SPEye to automatically extract protocol data from RPs in desktop web browsers; addressing the parallel but unique problem in the mobile ecosystem and with mobile apps as RPs is equally valuable but beyond our current scope.
\newline

\noindent
\textbf{SPEye-like tool to inform security decisions.}
\label{sec.discussion.securityextension}
SPEye demonstrates the feasibility of extracting SSO protocol data from RPs; it could be expanded to convey information about potential security weaknesses in addition to privacy information, and warn users about risks prior to committing to login decisions.
Security-related information extracted using methods from previous security tools can be presented to users in an SPEye-like workflow (but now showing security rather than privacy-related info).
For example, a security scan could use the data available through SPEye's existing framework to scan RP client-side code, e.g., for use of OAuth implicit flows (known to be weak) or lack of CSRF protection (as applied through the OAuth \texttt{state} parameter).
Such an augmented version of SPEye could warn users about vulnerable RP implementations before they commit to SSO login decisions.

\section{Related Work}

\noindent\textbf{SSO privacy tools.}
Existing tools used in privacy measurements primarily rely on browser automation to scan RP sites and execute OAuth and OpenID flows.
J\"arpehult et al.~\cite{jarpehult2022longitudinal} developed a Selenium-based tool to track differences in RP permission requests and IdP usage over a nine-year period.
The tool launches browser instances controlled by Selenium to crawl and collect data from RP implementations.
Similarly, Jannett et al.~\cite{jannett2024ssomonitor} built SSO-Monitor using Selenium for analyzing the security of SSO login workflows with three IdPs. They developed a visual-based approach to identify SSO login buttons, which allows identification of SSO options independent of the site's language or region.
Morkonda et al.~\cite{morkonda2021empirical} developed OAuthScope based on Selenium to extract differences in permission requests among SSO login options offered in RP sites.
Their tool supports data collection from RP sites across five different countries. 
OpenWPM~\cite{englehardt2016online} is a comprehensive privacy measurement framework that relies on Selenium to instrument user tracking techniques employed by site operators, including RP sites.
SPEye's approach differs from these tools as our end-user focus results in constraints (Sec.~\ref{sec.tool.requirements}) leading our design to explicitly avoid browser automation.
Instead, the SPEye extension analyzes loaded IdP and RP HTML pages to extract SSO protocol data and provide privacy information for end-users.

Increasing concerns about SSO privacy issues have motivated researchers to explore more privacy-friendly SSO protocols.
Fett et al.~\cite{fett2015spresso} proposed SPRESSO, an OAuth client-side forwarder that masks the RP's identity from the IdP by acting as the intermediary.
Hammann et al.~\cite{hammann2020privacy} proposed POIDC, a protocol extension to OIDC to hide RP identity from the IdP by masking the RP's client ID using cryptographic hash values.
Dey and Weis~\cite{dey2010pseudoid} proposed PseudoID to enable users to login to RPs using pseudonyms.
Their proposal uses a blind signature service to mask the user's real (online) identity.
Xu et al.~\cite{xu2023miso} proposed MISO to hide the RP's identity by introducing a middle component hosted in a trusted execution environment (TEE).
These existing proposals that aim to hide the RP's identity from the IdP involve protocol changes or modifications to the RP and/or IdP software which hinder their adoption.
Our work herein complements these privacy-enhancing proposals by improving the transparency of SSO permissions.

Shehab et al.~\cite{shehab2011roauth} proposed ROAuth as an extension to OAuth 2.0 and built a Firefox browser extension to allow Facebook users to configure (and possibly limit) requested permissions in SSO logins with Facebook.
Li et al.~\cite{li2019oauthguard} developed OAuthGuard, a Chrome extension for security analysis of RPs with Google SSO login.
Their extension monitors OAuth traffic to detect threats (e.g., CSRF attempts) then block or notify users about these vulnerabilities.
While the above extensions limit their analysis to one IdP, our tool covers three major IdPs because the primary goal of SPEye is to enable users to compare different login choices.
\newline

\noindent\textbf{Automated SSO security testing.}
Many security tools have been built to automate the testing of SSO implementations.
Zhou and Evans~\cite{zhou2014ssoscan} built SSOScan to test the security of Facebook SSO implementation in RPs by automating OAuth login flows and monitoring for certain known OAuth-related vulnerabilities such as credential leaks.
Sun and Beznosov~\cite{sun2012devil} performed a security analysis of OAuth 2.0 implementations across three IdPs and 96 RPs.
Their automated tracing of OAuth access tokens helped uncover new attack vectors including exploit opportunities for impersonation attacks.
Drakonakis et al.~\cite{drakonakis2020cookie} built XDriver to scan web login implementations (including RPs that offer login with Facebook and Google) for authentication and authorization flaws related to exposed cookies.
Ghasemisharif et al.~\cite{ghasemisharif2018single} reviewed access revocation in RPs following an IdP account compromise and presented the resulting security implications.
For example, they highlight the limitations of current SSO systems where it is difficult for users to revoke access tokens which is especially important if there is an attack on their account.
Similar to SPEye (Comparative mode), several tools (e.g.,~\cite{ghasemisharif2018single, zhou2014ssoscan, drakonakis2020cookie, morkonda2021empirical}) rely on custom heuristics to search RP HTML pages for CSS attributes and forms related to SSO login buttons.
These empirical studies also reveal inadequate security protections in many SSO systems, raising security and privacy concerns about users' online identity and personal information.
The privacy issues discussed in Section~\ref{sec.privacy.issues} complement prior work on OAuth security implications from identified vulnerabilities.
\newline

\noindent\textbf{SSO analysis approaches.}
Prior SSO research has explored various security analysis approaches to detect vulnerabilities in SSO implementations and protocols.
Mainka et al.~\cite{mainka2017sok} categorized SSO testing approaches used by previous research pointing out benefits and limitations, and built PrOfESSOS, a security analysis tool for testing OIDC implementations.
They recommend a testing approach that introduces a malicious IdP; given that IdPs are involved throughout the SSO workflow, this approach covers a larger attack surface.
Fett et al.~\cite{fett2016comprehensive} used formal analysis to evaluate and formally prove the security properties of OAuth 2.0.
To test compliance with security best practices, Rahat et al. built Cerberus~\cite{rahat2022cerberus} for web OAuth implementations, and OAuthLint~\cite{al2019oauthlint} for RP mobile apps.
Both these tools use static analysis to identify deviations from the OAuth 2.0 protocol. 
Yang et al.~\cite{yang2016model} built OAuthTester, a fuzzing tool to test the security of OAuth RP and IdP implementations through modification of protocol parameters such as OAuth 2.0 \texttt{state} and \texttt{redirect URI}.
Bai et al.~\cite{bai2013authscan} developed AuthScan to model RP implementations and automatically extract protocol specifications from SSO systems.
Using formal verification techniques, Wang et al.~\cite{wang2013explicating} modelled security assumptions made by OAuth app developers in the use of IdP SDKs.
These security testing approaches help find implementation errors and security flaws that impact SSO users' security. In contrast, our work focuses on the privacy implications of deliberate design choices made by RPs and IdPs.
\newline

\noindent\textbf{SSO privacy and usability.}
A survey on Google SSO by Balash et al.~\cite{balash2022security} found user concerns on granting access to sensitive data such as personal emails and contacts. They also find considerable variations in users' understanding of the necessity of specific permission requests, suggesting lack of transparency in current SSO systems.
A user study by Sun et al.~\cite{sun2011makes} found similar concerns in granting RPs access to personal user data.
A survey by Robinson and Bonneau~\cite{robinson2014cognitive} found the permission requests in Facebook SSO prompts ineffective in informing users.
Alaca and van Oorschot~\cite{alaca2020comparative} compared 14 web SSO systems on protocol features including security, privacy, and usability.
\newline

\noindent\textbf{Supporting privacy-informed decisions in non-SSO contexts.}
Previous research has focused on supporting privacy-informed decisions in non-SSO contexts such as mobile app permission systems.
The AppCensus tool, by Reardon et al.~\cite{reardon201950}, automatically instruments Android apps to identify their privacy behaviors and generate privacy labels.
For informing users during mobile app installation, Kelley et al.~\cite{kelley2013privacy} proposed displaying the permissions requested by an app before users commit to install the app.
This approach resulted in users choosing apps that requested fewer permissions, thus demonstrating the usefulness of presenting permissions to users before they make privacy-related decisions.
SPEye similarly prioritizes informing users before they make privacy decisions, by displaying web SSO permissions before users make login decisions.

Some researchers have proposed personalized privacy recommendations for supporting users' privacy choices.
Wijesekera et al.~\cite{wijesekera2017feasibility} built a permissions tool to automatically respond to mobile app permission requests based on inferences about the user's privacy preferences from past decisions.
Harbach et al.~\cite{harbach2014using} designed a personalized mobile app permission dialog that supplements permission requests with data samples from the user's data on the device.
A similar approach by Harkous et al.~\cite{harkous2016curious} examined third-party add-ons within websites, and proposed supplementing permission requests with personal insights (such as the user's interests or habits) that could be inferred from allowed data access.
While personalizing permission requests may be useful in certain contexts, it may not be viable in web SSO systems as the user data, which is released by the IdP directly to the RP, is by design unavailable until the user has already approved the requested access.

Mobile app permissions often require users to make a trade-off and simply give up privacy in order to use an app.
In contrast, many RP websites allow users to choose alternate login options---although these login choices and the related permissions are not always easily apparent---that reveal less user data while still allowing login (e.g., through a non-SSO login option).
SPEye supports web SSO privacy decisions by making the permissions more visible to users.
\newline

\noindent\textbf{Consent interface usability.}
Previous research has explored approaches to improve consent interfaces which often do not support privacy-informed decisions~\cite{cranor2023metrics}.
For example, many cookie consent interfaces use deceptive patterns to nudge users towards accepting cookies while making it harder to opt-out of cookies~\cite{utz2019uninformed}.
Habib et al.~\cite{habib2022okay} recommended improving consent interfaces by providing users information about available privacy choices and enabling users to make privacy choices with minimal effort.
In addition, standardized privacy dashboards (e.g., that show opt-out controls in a standard location) make it easier for users to indicate privacy preferences~\cite{habib2020its}.
Kelley et al.~\cite{kelley2010standardizing} introduced ``privacy nutrition labels'' to inform web users about privacy policies in a concise and standard manner.
Apple subsequently introduced similar privacy labels for iOS apps~\cite{applePrivacyLabels}.
SPEye uses a standard interface approach for informing users about SSO login choices and for enabling users to opt-out of permissions.

\section{Discussion}
Addressing SSO privacy issues such as the lack of transparency in permissions (Sec.~\ref{sec.privacy.issues}) is challenging because of the misaligned interests of different stakeholders.
It is difficult to incentivize RPs to promote privacy-friendly choices (e.g., to request less user data) as it may not be in the RP organization's business interest~\cite{bohme2007viability}.
On the other hand, the IdP might also have its own agenda (e.g., encouraging adoption of its services, or prioritizing ease-of-use over privacy) that may not align with that of privacy-conscious users or the RP.
With some RPs and IdPs, users are given limited information about particular IdP choices.
For example, as of June 2024, \url{nytimes.com}'s landing page shows a Google SSO  dialog that prompts the user to login using Google and indicates that the user's name, email address and profile picture will be shared with the RP site.
However, this prompt does not sufficiently inform the user about the available login choices because: (i) the prompt is shown on the RP's \textit{landing} page which does not show the other login options (i.e., login using Facebook, Apple, or email address and password), which are displayed only on the RP's (secondary) \textit{login} page; and (ii) the prompt only describes the information revealed when choosing Google SSO,  and it does not inform about other login options that reveal less information (e.g., the non-SSO login does not reveal name or profile picture).
This type of interface preferentially provides higher visibility of one IdP over others, possibly nudging users towards this choice.
We argue that in the absence of a tool such as SPEye providing relevant information for users to make informed decisions, RPs and IdPs may choose to prioritize their own business interest over user privacy goals.

The remainder of this section discusses SSO privacy issues related to SPEye and how similar tools can help users make informed privacy choices. We also suggest a list of changes for stakeholders (including RPs, IdPs, and OAuth specification designers) to consider in order to improve the transparency and the accountability of SSO systems.

\subsection{SSO privacy considerations}
\label{sec.discussion.otherconsiderations}

\noindent\textbf{Metadata exposure to IdPs.}
SSO protocols involve RP requests to the IdP to get access tokens and SDK scripts (in the SDK pattern RPs, as described in Sec.~\ref{sec.code.patterns}).
Consequently, the IdP can see the user's visit to the RP site (through OAuth redirect URI or HTTP referrer), including metadata such as timestamps for the visit, which raises privacy concerns relating to IdP tracking of the user's browsing activity~\cite{xu2023miso}.

To inform users about SSO permissions with each SSO login option, SPEye's Comparative mode makes a request to each IdP (described in Sec.~\ref{sec.tool.extension}).
These requests reveal site visit metadata to each of three IdPs, as opposed to the one IdP the user might select to login. 
Users may view this one-time release of metadata as acceptable assuming that SPEye is primarily used during signup when selecting a login on a new RP site, not for each subsequent RP visit. Alternatively, users can limit release of metadata to a single IdP by using the Focused mode.
However, note that site visit metadata equivalent to that released by Comparative mode is often already available to IdPs through embedded third-party services in the background, whether users are aware or not, as explained next.

Web sites (including RP sites) typically embed third-party scripts for advertising, analytics, or social network plugins~\cite{mayer2012third}.
A study across the top 1-million sites found that the most popular third-party services were Google and Facebook services~\cite{englehardt2016online}. For example, \texttt{google-analytics.com} itself was found in about 65\% of the sites, not counting other Google-owned services.
The third-party prevalence of Google and Facebook, along with RP requests (in the SDK pattern) to get SDKs from IdPs including Apple, mean that the above privacy issue relating to tracking of browsing activity by third parties exists regardless of SPEye's IdP requests.

In this work, we did not explore users' mental models related to privacy concerns about IdPs collecting metadata about websites visited, as distinct from IdPs releasing user data to RPs.
We also note that SPEye does not collect or retain any user data (and thus does not raise any privacy compliance issues, e.g., related to GDPR).
\newline

\noindent\textbf{Permissions requested after first login.}
SPEye enables comparison of permissions requested by an RP at the initial IdP login prompt where permissions are not usually visible to users, but of course cannot include data the RP might request at a later time (e.g., after first login).
For example, an RP that requests only the user's name from the IdP with whom the user has completed login might at a later point (in an RP UI, as Fig.~\ref{figRakutenError}) ask the user to enter their email address.
A privacy-unfriendly RP might request additional permissions gradually at various points in its workflow (escaping SPEye), perhaps across many different interactions with the RP, leading to a form of `permission creep' (a progressive increase in permissions) as the user performs different tasks.
The UI workflow for these extra requests explicitly prompts the user to grant or deny the additional permissions `just-in-time', without having to complete login or make a new login choice.

IdP guidelines (e.g.,~\cite{facebookOauthPermissions}) suggest that RPs should request additional permissions in the context where the data is needed.
Although this design aims to give users better control over what data is requested and how it might be used, it withholds from users overall context about how much other data is already being collected or how this new permission fits into this overall picture. 
In a type of deceptive pattern~\cite{narayanan2020dark}, an RP could request minimal information during the initial login (and thus appear privacy-friendly in SPEye) and then gradually request additional permissions, hoping that the user will not realize the extent of data collected, or will feel compelled to stick with their decision (e.g., due to inconvenience or impossibility of transferring an existing RP user-account to a different IdP).

SPEye provides comparative information necessary to make an informed initial login choice and offers a model of the type of information that helps users make informed decisions. Future research may explore how to keep users informed throughout their relationship with an RP.
\newline

\noindent
\textbf{RP handling of opted-out permissions.}
Except for basic information such as the user's name and profile picture, major IdPs (e.g., Facebook) classify other permissions as `optional' and allow users to decline such permissions requested by an RP. The apparent intent by the IdPs is to give users control over their data.
However, as described in Sec.~\ref{sec.privacy.issues} and Fig.~\ref{figRakutenError}, some RPs use tactics to persuade users to grant permissions despite being labelled as optional on the IdP's login prompts (if users want to use the given IdP).
Although RPs may have legitimate uses for the data, such workarounds suggest the potential presence of a misleading or deceptive pattern (cf. above) in RP implementations, bypassing any intended IdP  privacy controls.
SPEye is unable to detect such RP workarounds because they occur outside of the SSO login interaction.
Conflicting cues in existing SSO UIs, such as whether access to specific user data is mandatory (or not) for successful login at the RP site, are visible in Fig.~\ref{figRakutenFB} and ~\ref{figRakutenError}. These conflicting cues may confuse users and consequently reduce user trust in the privacy of SSO systems.
Tools like SPEye may help users better manage access control to data items by offering opt-out options at the start of IdP login workflows.

\subsection{Potential Changes for Stakeholders}
\label{sec.recommendations}
To improve the privacy and the transparency of web SSO systems, we suggest the following protocol and interface changes that could be adopted by different stakeholders.

\begin{itemize}
\item [C1:] 
\textit{When registering with an IdP,
an RP could be required to state descriptions committing to the intended use of  
each OAuth user-data attribute they plan to request from users.}
\end{itemize}

\noindent
This might be accompanied by  \textit{value labels} (cf.\ \textit{privacy nutrition labels}\footnote{\url{https://cups.cs.cmu.edu/privacyLabel/}})
that convey intended uses and benefits (if any) to users, possibly with cooperation from the browser vendors. 
The IdP/browser app could show these labels in a standardized UI before asking users to authorize release of RP-requested data attributes.
In addition, browser vendors could promote privacy-informed choices by adopting SPEye-like UIs to enable comparisons of different SSO privacy choices.

\begin{itemize}
\item [C2:] \textit{The OAuth specification could be revised to allow
optional scope parameters distinct from those denoted mandatory for RP operation.}  
\end{itemize}

\noindent
Although it is currently already possible for IdPs to allow users to opt-out of scope parameters, IdPs currently cannot distinguish which parameters are mandatory for RP operation.
RPs and IdPs that favor transparency could use an \textit{optional} scope parameter to denote user data attributes that are not required for an RP to provision its core features (this might be compared to necessary vs. optional cookies in many cookie consent interfaces).
RPs could then provide the value labels (as mentioned above) to convey the benefits for users opting-in to the optional permissions.

The above changes could allow audits or privacy compliance checks (by IdP or third parties), and support informed choices by privacy-conscious users.
Over time, this could result in RP-IdP pairs following the privacy best practice of requesting only \textit{need-to-know} data, and privacy-friendly IdPs gaining a ``preferred-IdP'' status.

\begin{itemize}
\item [C3:] \textit{Reputation-based or data-driven community efforts could provide
privacy ratings for SSO login options at popular RPs.}
\end{itemize}

\noindent
Privacy ratings as suggested in change C3---which would complement SPEye---may help privacy-conscious users make informed decisions and help inform the community about RP privacy practices.
Such ratings may also motivate RP organizations to limit the amount of personal data they collect. Separate from SSO, Mozilla has published similar privacy ratings\footnote{\url{https://foundation.mozilla.org/en/privacynotincluded/}} of several popular consumer devices and apps to highlight concerning privacy and security practices.

Many of these suggestions may of course face hurdles, one (as noted earlier) being that the agendas of commercial RPs, IdPs, and browser vendors are often not aligned with those of privacy-conscious users.
However, individual organizations within a stakeholder group might have their own business models and interests, e.g., compared to Facebook or Google, Apple's SSO platform offers significantly fewer user data attributes to RPs~\cite{morkonda2021empirical} and does not rely as heavily on advertising revenue.
Thus, some privacy-oriented IdPs and browser vendors (e.g., DuckDuckGo, Firefox) may be inclined to adopt these changes to improve privacy, e.g., to increase their market share among privacy-conscious users.
We also argue that awareness and discussion of technical possibilities are important steps towards supporting user privacy and towards shedding light on any RP deceptive patterns~\cite{narayanan2020dark}.

\section{Concluding remarks}
Current OAuth-based web SSO workflows do not support privacy-informed login decisions---users would have to complete authentication with each SSO login option in order to be able to view and compare IdP permissions on a given RP site.
Due to this lack of transparency, users risk unintentionally making choices revealing more information than desired.
The SPEye Chrome extension addresses this lack of transparency by extracting from RP and IdP sites information to inform users' SSO login decisions by enabling a real-time comparison of permission requests across multiple IdPs.

We contextualized SPEye's two main workflow approaches (Focused and Comparative) through the identification of four RP client-side code patterns, highlighting similarities and variations across RP implementations.
SPEye's design enables its Focused mode to extract permission information in real time across all RPs, from the login pages of the IdPs that the current tool targets; its Comparative mode presents a real-time permission comparison of SSO login options at the start of the user's login flow for a subset of RP login pages.

Beyond informing the design of SPEye, we believe these patterns can motivate new web SSO tools that use the same code features as SPEye to extract protocol data from RP implementations, e.g., to inform users in real time about RP security weaknesses. 
SPEye focused on privacy; tools that investigate and report on the privacy practices of service providers help raise public awareness of privacy issues, motivating transparent privacy practices.
Our work on SPEye takes a step towards increasing transparency in web SSO systems by enabling users to be informed about permissions and ``Sign in with (more) privacy''.

\begin{acks}
    We thank the anonymous referees for helpful reviews. Morkonda acknowledges a scholarship from OGS. Chiasson acknowledges NSERC for funding of an Arthur B. McDonald Fellowship and a Discovery Grant. Van Oorschot acknowledges funding from NSERC for a Discovery Grant. We thank the members of the CCSL, CISL and CHORUS labs at Carleton University for their valuable feedback.
\end{acks}

\bibliographystyle{abbrv}
\bibliography{references}

\appendix

\section{SPEye implementation details}
\label{sec.app.identifying.sso.components}
This section provides details on SPEye's implementation related to our analysis in Sec.~\ref{sec.tool}.

SPEye identifies SSO login options in RP sites using CSS Selectors to search the DOM for strings such as ``Sign in with'' and related elements.
We built these Selectors based on strings (converted to lowercase for comparison) and associated elements found in 21 RP sites in our training set, as described in Sec.~\ref{sec.tool.testing}.
Table~\ref{tableSsoSelectors} shows the distribution of the SSO strings we found in the 36 HTML-based SSO implementation.

To identify IdP endpoints related to SSO login options (i.e., authorization servers), we used regular expressions (shown in Listing~\ref{code.sso.idpregex}) derived from IdP documentation.

\begin{table}[tb]
\centering
\caption{Number of RP sites that display SSO login options using the specific strings within HTML elements.}
\begin{tabular}{l|l|c}
\toprule
SSO string & DOM element & \# of RP sites \\
& & ($N=36$) \\
\midrule
sign in with & \texttt{span/text()} & 5 \\
sign in with & \texttt{div/text()} & 3 \\
sign in with & \texttt{a/text()} & 2 \\
sign in with & \texttt{small/text()} & 1 \\
sign in & \texttt{button/text()} & 2 \\
\midrule

continue with & \texttt{span/text()} & 3 \\
continue with & \texttt{div/text()} & 3 \\
continue with & \texttt{a/text()} & 3 \\
continue with & \texttt{button/text()} & 1 \\
\midrule

log in with & \texttt{span/text()} & 3 \\
log in with & \texttt{div/text()} & 1 \\
log in with & \texttt{p/text()} & 1 \\
login with & \texttt{a/text()} & 2 \\
login with & \texttt{p/text()} & 1 \\
\midrule

login via & \texttt{p/text()} & 1 \\
connect using & \texttt{span/@data-text} & 1 \\
or use & \texttt{span/text()} & 1 \\
idp link & \texttt{a/@title} & 1 \\
idp link & \texttt{iframe/@src} & 1 \\
\bottomrule
\end{tabular}
\label{tableSsoSelectors}
\end{table}

\begin{listing}[b]
\begin{minted}[frame=single]{html}
https://(.*)\\.facebook\\.com/login(.*)
https://(.*)\\.facebook\\.com/oauth(.*)
https://graph\\.facebook\\.com/(.*)
https://(.*)\\.facebook\\.com/(.*)/oauth(.*)
https://(.*)\\.google\\.com/(.*)/oauth(.*)
https://oauth2\\.googleapis\\.com/(.*)
https://openidconnect\\.googleapis\\.com/(.*)
https://googleapis\\.com/oauth(.*)
https://(.*)\\.apple\\.com/auth(.*)
\end{minted}
\caption{Regular expressions for identifying IdP endpoints.}
\label{code.sso.idpregex}
\end{listing}

\begin{figure}[b]
 \centering
    \frame{\includegraphics[width=0.45\linewidth]{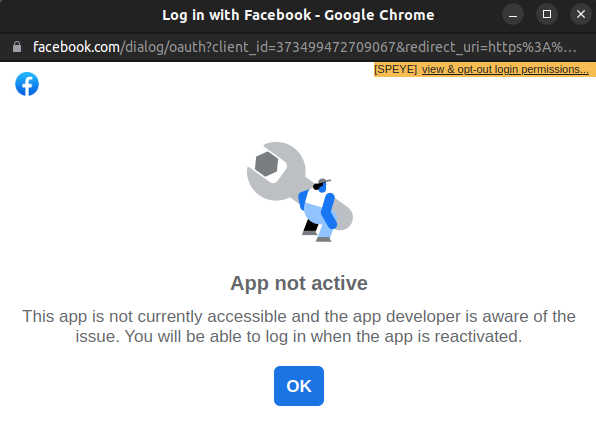}}
    \caption{A temporarily blocked message displayed on Facebook prompt when attempting to log in to \url{myspace.com} using Facebook.}
    \label{figFbRpBlocked}
\end{figure}

\section{Recent Facebook SSO policy changes}
\label{sec.app.fbchanges}
In February 2023, we noticed changes in Facebook's SSO platform policy~\cite{Fb2023NewRulesForAdvancedAccess} that blocked some RP sites from requesting permissions other than name and email address.
We describe these changes and implications to our work in Sec.~\ref{sec.tool.idp.deviation}.

In a subset of the RPs we observed, Facebook either removed permissions other than name and email address from the authorization requests received from the RP, or completely blocked the specific RP's users from using Facebook to log in to the RP site.
Fig.~\ref{figFbRpBlocked} provides a screenshot of an RP where login requests to Facebook were temporarily blocked.

\end{document}